\documentstyle[psfig]{l-aa}

\begin{document}

   \thesaurus{11         %A&A Section 11: Galaxies
              (04.01.2;  %Atlases
               04.03.1;  %Catalogs
               11.07.1;  %Galaxies: general
               11.04.1;  %Galaxies: distances and redshifts
               11.06.2;  %Galaxies: fundamental parameters
               11.14.1;  %Galaxies: nuclei
               11.19.7)} %Galaxies: statistics
   \title{A statistical study of the spectra of very luminous IRAS galaxies}

   \subtitle{I. Data}

   \author{H. Wu\inst{1}, Z.L. Zou\inst{1}, X.Y. Xia\inst{2} and Z.G. Deng
\inst{3} }

   \offprints{H. Wu}

   \institute{Beijing Astronomical Observatory, 
Chinese Academy of Sciences, Beijing 100080, China
\and Department of Physics, Tianjin Normal University, Tianjin 300074, China
\and Department of Physics, Chinese Academy of Sciences Graduate School, Beijing 100039,  China
             }

   \date{Received ; accepted }

   \maketitle

   \begin{abstract}\footnotemark{}
    \footnotetext{Table 2-4 and Figure 4,9 are only available in electronic 
form at the CDS via anonymous ftp to cdsarc.u-strasbg.fr(130.\\
79.128.5) or via http: 
//cdsweb.u-strasbg.fr/Abstract.html.}

 This paper presents the results of spectral observations for the largest complete sample 
of very luminous IRAS galaxies obtained to date. The sample consists of those 73
objects for which
$\rm log(L_{IR}/L_{\odot}) \geq 11.5 $ ($\rm H_{0}=50 km s^{-1} Mpc^{-1}$, $\rm q_{0}=0.5$) and $\rm mag \leq 15.5$, and was extracted from the
2 Jy IRAS redshift catalog.
 All the spectra were obtained using 2.16m telescope of Beijing Astronomical 
Observatory during the years 1994-1996. A total of 123 galaxy spectra were obtained with spectral ranges of 
4400$\rm \AA$ to 7100$\rm \AA$ and 3500$\rm \AA$ to 8100$\rm \AA$
at resolutions of 11.2$\rm \AA$ and 9.3$\rm \AA$ respectively. In addition to the 73 spectra for sample galaxies, we also present
 spectra for ten non-sample galaxies and a further 40 for the companions of sample galaxies.
The data presented include nuclear spectrum and the parameters describing the emission lines, absorption lines and continua as well as
DSS images and environmental parameters.

   \end{abstract}

\section{ Introduction}
  The Infrared Astronomical Satellite(IRAS) all-sky survey provided a large 
database(IRAS Point Source Catalog, Version 2, 1988) of infrared galaxies. 
Based on recently completed redshift surveys of IRAS
galaxies, statistical spectroscopic analysis of very luminous IRAS galaxy samples can now be performed.
Sanders et al. (1988) have already studied the spectra of a small sample of ten ultraluminous
IRAS galaxies selected from IRAS Bright Galaxy Survey (BGS) of Soifer et al.(1986). 
Recently Kim et al. (1995) completed a spectroscopic survey based on a large sample of luminous IRAS galaxies that was extracted from both the BGS  
and infrared warmed catalogs. However, the infrared luminosity range of Kim et al.'s sample is 
from $\rm 10^{10.5} L_{\odot}$ to $\rm 10^{12.5} L_{\odot}$ and it 
therefore can not represent very luminous IRAS galaxies.

  This paper presents the spectroscopic data for a large complete sample of very luminous IRAS galaxies (VLIRGs), 
extracted  from 1.936 Jy Redshift Survey of Strauss et al.(1992). The DSS images
are also presented, here. A detailed analysis of the spectra and environment will be presented in paper II (Wu et al. 1997).

\section{ Sample selection}
    
  Strauss et al. (1990) selected 5014 objects from the IRAS database(IRAS Point Source Catalog, Version 2, 1988) according to the
criteria: 
\begin{itemize}
\begin{enumerate}
\item  $\rm f_{60} > 1.936 Jy$ 

\item  $\rm f_{60}^{2} > f_{12} f_{25}$ \\
a color criterion distinguishing galaxies from objects in the Galaxy

\item  Galactic latitude $\rm |b| > 5^{\circ}$

\end{enumerate}
\end{itemize}
Strauss et al.(1992) then went on to publish the redshifts of the 2658
objects which were galaxies (here after the 2Jy samples).

  Our VLIRG sample is a subset of the 2Jy sample. Considering the observatory site, instruments and  
possible observation times, we selected galaxies according to the following criteria:
\begin{itemize}
\begin{enumerate}
\item $\rm \delta \geq  0^{\circ}$ \\

\item $\rm log(L_{IR}/L_{\odot}) \geq 11.5$ \\
         ($\rm H_{0}=50 km s^{-1} Mpc^{-1}, q_{0}=0.5)$\\

\item $\rm  mag \leq 15.5$ \\

\end{enumerate}
\end{itemize}

  Here the magnitudes are from the 2Jy catalog. These criteria guarantee S/N ratios good enough for spectroscopic classification.
73 of 2Jy-catalogue VLIRGs met these criteria.

  On account of the low S/N at 12$\mu$m and 25$\mu$m for some of sources in 2 Jy sample,
we used Dennefeld et al.'s (1986) formula for calculating the infrared flux,
between 42.5$\mu$m and 122.5$\mu$m:
\begin{equation}
 {\rm   F_{IR}=1.75[2.55S_{60} + 1.01 S_{100}] 10^{-14} \rm  W m^{-2} }
\end{equation}
where, $\rm S_{60}$ and $\rm S_{100}$ are the fluxes at 60$\mu$m and 100$\mu$m respectively.

The infrared luminosity  is therefore:
\begin{equation}
 \rm    L_{IR}=4\pi D^{2} F_{IR} 
\end{equation} 
for each object,where D is the distance from the Galactic center. 

\section{The sources lists and general properties}

  Table 1 lists the 83 target galaxies, of which 73 belong to the complete sample. Of the other ten objects in brackets,
four of them have slightly lower infrared luminosities than $10^{11.5}$ $\rm L_{\odot}$ and six
are fainter than our magnitude limit ( $\rm mag >  15.5$ ). This table presents the 
infrared luminosities in unit of $\rm L_{\odot}$, magnitudes
and redshifts. All of these data were derived from the 2Jy catalog, 
except that in the case of IR09517+6954 (M82). We can not use redshift as a distance indicator on account of the proximity.
For this reason, we adopted the distance value given by Tully (1988).
These objects flagged with asterisks in column 1 are the sources which are also included  in 
the IRAS Bright Galaxy Sample ($\rm S_{60}$ $> 5.24 \rm Jy$) of  Kim et al.(1995). A detailed comparison between these two
samples will be made in paper II.

  The spatial distribution of the 73 sample galaxies on the sky is shown in Fig.1. The dotted curve is the celestial
equator. The scarcity of objects near galactic declination $\rm b \sim 0^{\circ}$ is due to the 
matching Zwicky catalog (1961-1968) which misses galaxies at low latitude.

\begin{figure}
\centerline{\psfig{figure=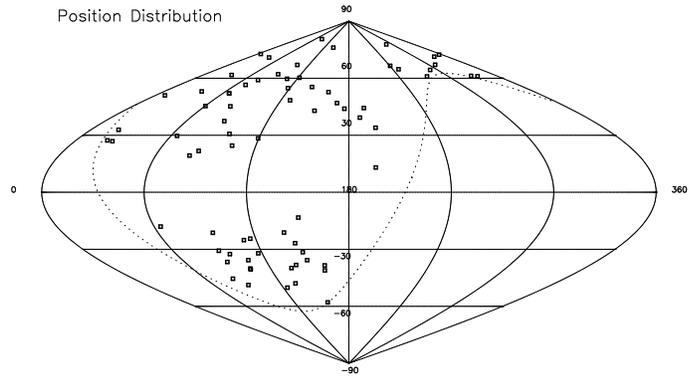,height=5.0cm,angle=270}}
      \caption{The distribution of very luminous IRAS galaxies in the sky shown in equal area projection using  Galactic coordinates.
              }
         \label{Fig1}
    \end{figure}

\begin{figure}
\centerline{\psfig{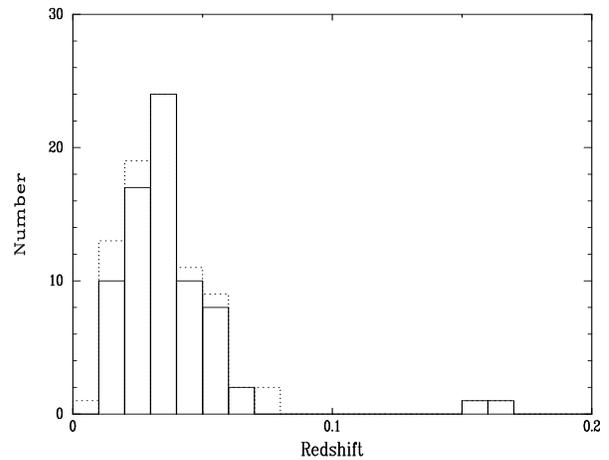}}
      \caption{Distribution of redshifts. The solid boxes represent our
            complete samples while the dotted boxes include the other
            ten galaxies.
              }
         \label{Fig2}
    \end{figure}

\begin{figure}
\centerline{\psfig{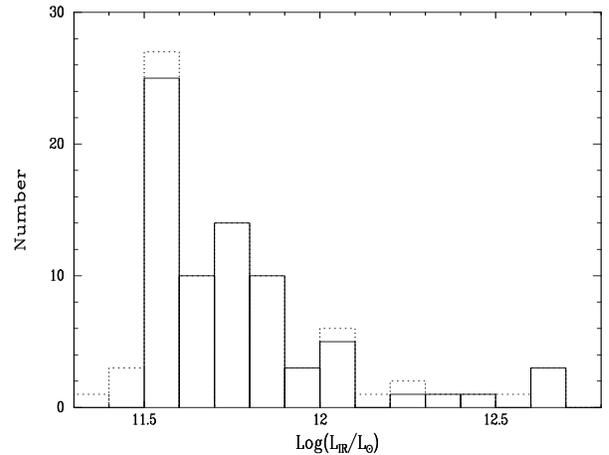}}
     \caption{Distribution of infrared luminosities. The 
            solid and dotted boxes represent the same galaxies
	    as they do in Fig.2.
              }
         \label{Fig3}
    \end{figure}

  Fig.2 shows the sample distribution as a function of redshift. The solid boxes represent  the complete
sample, while the dotted boxes include the other ten  sources. It is clear that the
redshifts concentrate in the range 0.02 to 0.04. The
median value is 0.0324 ( corresponding to a recession velocity $\rm  9700 \rm kms^{-1}$) which is larger than
the value $5900 \rm km s^{-1}$ obtained by Kim et al.(1995) for their sample.

  The infrared luminosity distribution is shown in Fig.3. The solid and dotted
boxes correspond to the same samples as in Fig.2. The counts decrease rapidly as the
infrared luminosity increases. Most of the galaxies have infrared luminosities
$\rm log(L_{IR}/L_{\odot}$) between 11.5 and 12.0.

\section{Spectroscopic observations}

  The spectroscopic observations were made between Apr.15, 1994 and 
Feb.15, 1996. We obtained a total of 123 galaxy spectra.
All of the observations were made with 2.16m telescope at the Xinglong Station of
Beijing Astronomical Observatory using Zeiss universal spectrograph with a grating of 195$\rm \AA$/mm dispersion.

  Before Nov.5, 1994, a Tek 512x512 CCD which covered a 2700$\rm \AA$ range from 4400$\rm \AA$
to 7100$\rm \AA$ 
at a resolution of 11.2 $\rm \AA$ (2 pixels) was used. After that, a Tek 1024x1024 CCD
which covered a 4600$\rm \AA$ range from 3500$\rm \AA$ to 8100$\rm \AA$ at a resolution of 9.3$\rm \AA$ (2 pixels) was employed.

  In most cases, slit width of about 3" was chosen to match the typical seeing disc 
at Xinglong Station, but occasionally, the seeing disc was smaller than 2" or larger than 5". 
Thus may affect the spectral classification. Slit position angles of 
$90^{o}$ were generally used. 
For objects with two close nuclei ( separation $\leq$ 2'), the slit was rotated 
such that two spectra could be obtained simultaneously.

  The seeing was about 3" to 4" on most of the observation nights. In order to perform 
a relative flux calibration,  KPNO standard stars were also observed on each night.

 Table 2 lists the observations by target objects. The standard IRAS name , together with one upper case letters representing 
the individual components( See Fig.9 for identifications) are listed in Column 1. Column 2 to Column 10
give for each source:  coordinates, observation date
(Beijing Time), exposure time, airmass, slit width and position angle respectively.
The sources flagged with ticks in Column 11 were observed during good
weather condition.

\section{Data reduction }

  All of the data reductions were performed using IRAF
\footnotemark{} \footnotetext{IRAF is provided by NOAO}.
The IRAF packages
CCDRED, TWODSPEC and ONEDSPEC were used to reduce the long-slit spectral
data. The CCD reductions included bias subtraction, flat-field correction and
cosmic-ray removal. The dark counts were so low that their subtraction was not performed.
  We adopted Horne's (1986) avariance weighed algorithm in order to extract the spectra.
As for the aperture size, we adopted  similar method to that of Kim et 
al.(1995) to minimize the aperture-related effects on the nuclear spectra. The 
apertures were varied according to the redshift of the objects so that they approached
diameters of 2 kpc( $\rm H_{0}=50 km s^{-1} Mpc^{-1}$) for galaxies with $\rm z < 0.034$, but were fixed 
at 4" for objects at larger redshifts as the seeing was typically 3" to 4" and the pixel 
sizes was 1.3". The only two exceptions were for IR12323+1549A and IR09517+6954 (M82) which 
are quite nearby and a 2 kpc aperture was too large for them. 

   An Fe-He-Ar lamp was used for the wavelength calibrations. 
More than 20 lines were used to establish the wavelength scale by fitting a first-order spline3
function. The accuracy of the wavelength calibration was better than 1$\rm \AA$.

  On most nights, more than two KPNO standard stars were used to perform the relative flux calibration.
Atmospheric extinction was corrected using the mean 
extinction coefficients for the Xinglong station, that were measured out by BATC multi-color survey (Yan, 1995).  

  The telluric $\rm O_{2}$ absorption bands at 6870$\rm \AA$
and 7620$\rm \AA$ did heaver effect on those emission lines at the similar observed wavelength. 
We therefore constructed an artificial - the spectrum
of standard stars setting every wavelength to unity except these wavelength corresponding to  the $\rm O_{2}$ bands.
Division by the artificial spectrum could therefore removed telluric absorption bands.
In most cases, this technique worked well, especially for the band near 6870$\rm \AA$. However in
some cases, this correction seems not to have been very satisfactory and affected the measurement 
of the emission lines at 7620$\rm \AA$.

\section{Spectra obtained}

    Optical spectra are presented for all program objects in Fig.4.
They have all been corrected for telluric absorption, though some of 
corrections were not as good as we would have thought.

%\begin{figure}
%\centerline{\psfig{figure=f4_1.ps,height=25.0cm,width=15.0cm}}
%\caption{The spectra of VLIRGs f$_{\lambda}$ vs. $\lambda_{observed}$.
%         The units of the vertical axis are ergs s$^{-1}$ cm$^{-2}$
%         \AA$^{-1}$.
%         }
%\label{fig4}
%\end{figure}

\setcounter{figure}{4}
\begin{figure}
\centerline{\psfig{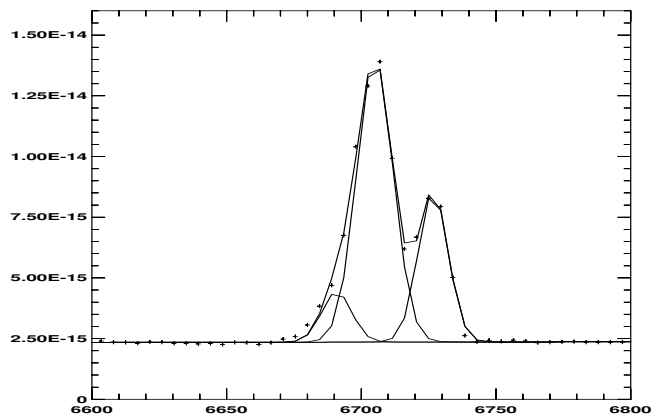}}
     \caption{Example of Gaussian fit for H$\rm \alpha$+[NII]$\lambda6584+\lambda6548$
            emission. Three narrow Gaussian components are used.
	    The pluses are observed data.
              }
         \label{Fig5}
    \end{figure}

\begin{figure}
\centerline{\psfig{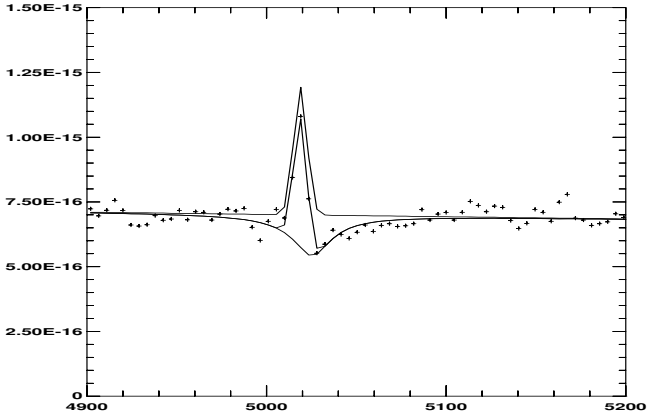}}
     \caption{Example of multi-component fit for the galaxies with 
            strong Balmer absorption. One Lorentz absorption and
	    one narrow Gaussian emission components are used.
              }
         \label{Fig6}
    \end{figure}

\begin{figure}
\centerline{\psfig{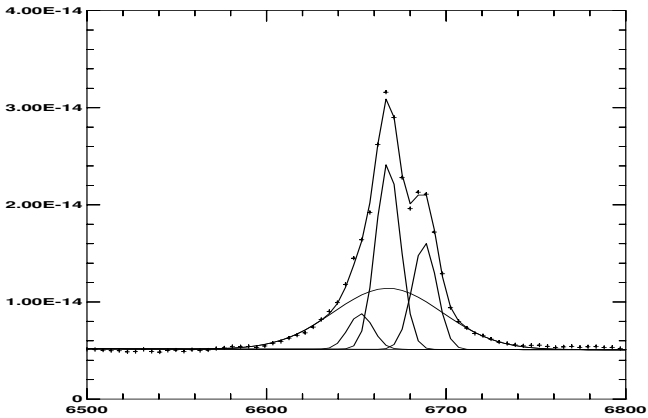}}
     \caption{Example of Gaussian fit for Seyfert 1s. One broad and
            three narrow Gaussian emission components are used.
              }
         \label{Fig7}
    \end{figure}

\begin{figure}
\centerline{\psfig{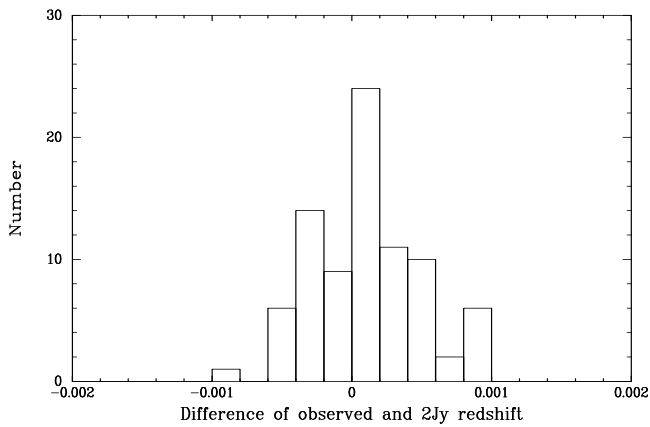}}
     \caption{Distribution of redshift differences between our
            data and the 2Jy redshift catalog. 
              }
         \label{Fig8}
    \end{figure}

   The measurements of emission line, absorption line and continuum strengths were performed within
the IRAF environment using tasks (SPLOT and SPECFIT). For isolated emission lines such as 
$\rm [OIII]\lambda5007$, $\rm [OII]\lambda3727$ and $\rm [OI]\lambda6300$, both Gaussian fitting and direct integral methods were used.
For the blended lines such as $\rm H{\alpha}$, $\rm [NII]\lambda6548,\lambda6584$, and $\rm [SII]\lambda6716,\lambda6731$ 
double lines, we employed a multi-Gaussian component method using task SPECFIT,
to deblend them. There are three parameters for each 
Gaussian component: the central wavelength, the total flux and width of the emission line; 
and two parameters for each continuum component: the flux and the slope. In 
order to speed up the convergence, some limit conditions were adopted. For example, 
we fixed the center wavelengths of several components relate to one another. In Fig.5, for example, we used the simplex
algorithm for the fitting and obtained the result via a chi-square 
minimization process. As for the spectra with obvious 
$\rm H{\beta}$ absorption indicating an underlying stellar population, a similar method was used. 
Because of the coexistence of $\rm H{\beta}$  emission and absorption, one emission and one 
absorption component were used for the $\rm H{\beta}$ fitting. The  
wide absorption wings due to stellar populations often cause the absorption to be overestimated 
if Gaussian model is adopted.
To solve this problem, we adopted the Lorentz model as shown in Fig.6, and the results 
seem better. Some of our sample galaxies are Seyfert-like, and their spectra  could not fitted well 
with only single Gaussian component for each of  $\rm H{\alpha}$ , $\rm H{\beta}$. In those case, we combined one 
narrow ($\rm < 1000 km s^{-1}$) and one broad ($\rm >1000kms^{-1}$) Gaussian component as shown in Fig.7.

    The relative emission-line fluxes are listed in Table 3 
for all objects. The typical uncertainty in these measurement 
 is about 10\%. Colons (:) and semicolons (;) indicate values with relative uncertainties 
at about 30\% and 50\% respectively. For the line $\rm [OII]\lambda3727$, we could not obtain a value with an uncertainty of less than 20\%, 
because it was at the blue end of the spectra, which can be affected by the 
low Q.E., lower S/N and poor flux calibration. 
In some cases, $\rm [OI]\lambda6300$ lines were heavily 
affected by the nearby emission lines $\rm [SIII]\lambda6312$ which could increase
the uncertainty. The telluric absorption bands 6870$\rm \AA$ and 7620$\rm \AA$ was 
also enlarge the uncertainty in the measurement of lines near them, despite
the corrections performed. The double lines $\rm [SII]\lambda6716,\lambda6733$, 
could sometimes be separated,
but when this was not possible, only the combined values are given. 

    The measured redshifts, observed $\rm H{\alpha}$ fluxes, 
equivalence widths of $\rm H{\alpha}$ emission lines, NaID absorption lines and $\rm H{\beta}$  absorption lines, and two
continuum fluxes (at $\rm 4861\AA$ and 6563$\rm \AA$) are listed in Table 4.  
The typical uncertainty in the measured $\rm H{\alpha}$ fluxes was 15\%, as the 
$H{\alpha}$ lines often had to be deblended from an overlapping [NII] emission line.

Finally, we compare our measured redshifts with those of corresponding sources in the 2Jy redshift sample. 
The distribution of redshift differences is plotted in Fig.8. The mean 
redshift difference is 0.000054 and scatter is 0.00038. This means that our measurements 
agree well with these of the 2Jy catalogue.

\section{DSS images}
  In  order to be able to study the relationship between nuclear activity and degree of interaction of VLIRG in a future paper in this series, 
we extracted the optical images for all our objects from
the CD-ROM version
of Digital Sky Survey\footnotemark{}\footnotetext{DSS CD-ROM is provided by 2.16m Telescope, Xinglong Station, Beijing
Astronomical Observatory}. The contour maps using IRAF tasks, we presented in Fig.9.

\section{Summary}
  We have presented the observation data of optical nuclear spectra and DSS
images of a sample of very luminous IRAS galaxies from 2Jy catalog. In the following
paper (Wu et al., 1997, Paper II), we will give the results of spectral analysis and 
environmental study, in the same time, address the possible model and evolutionary
sequence of very luminous IRAS galaxies.
%\begin{figure}
%\centerline{\psfig{figure=fig9.ps,height=25.0cm}}
%\caption{Contour maps of VLIRGs. Crosses are position of IRAS sources.
%         A,B,C,D are observed galaxies.
%         }
%\label{Fig9}
%\end{figure}

\begin{acknowledgements}

  The authors would like to thank Prof. J.Y. Hu and Dr. J.Y. Wei.
Their active cooperation enabled all of the observations to go through smoothly.
We also thank Mr. H.J. Yan for kindly providing the extinction data for Xinglong and
Prof. J.S. Chen and the rest of his group for help with the data reduction.
We are also grateful to Gerard Kriss kindly providing for
his procedure SPECFIT. 
Special Thanks to Dr. C. Young for his hard work of English revision of this paper.
Finally, we are most grateful to Dr. Michael Strauss for his valuable advice.
This work was partially supported by the Chinese National Science Foundation.

\end{acknowledgements}

%Table begin
\begin{table}[p]
\begin{tabular}{lccc|lccc}
\multicolumn{8}{l}{\large { \bf Table 1} { VLIRGs from IRAS 2Jy Catalog}}\\
\multicolumn{8}{l}{}\\
\hline
\hline
\small
IRAS$^{a}$  & $\rm Log(L_{IR}/L_{\odot})$  &   mag &  z & IRAS & $\rm Log(L_{IR}/L_{\odot})$ &   mag &  z \\
 (1)        & (2)       &   (3) &  (4) & (1) & (2) & (3) & (4)\\
\hline
00189+3748   & 11.572   & 15.30 & 0.0364  &13136+6223*  & 11.937   & 15.10 & 0.0311 \\
00267+3016   & 11.966   & 14.80 & 0.0504  &13183+3423*  & 11.863   & 14.80 & 0.0230 \\
00509+1225   & 11.772   & 14.00 & 0.0604  &13299+1121   & 11.516   & 14.50 & 0.0317 \\
01173+1405*  & 11.868   & 14.90 & 0.0312  &13362+4831*  & 11.706   & 14.10 & 0.0275 \\
01324+2138   & 11.629   & 15.30 & 0.0472  &13373+0105*  & 11.701   & 13.80 & 0.0225 \\
01484+2220*  & 11.851   & 13.70 & 0.0324  &13428+5608*  & 12.392   & 15.00 & 0.0373 \\
01572+0009   & 12.665   & 15.20 & 0.1630  &13458+1540   & 11.821   & 15.00 & 0.0570 \\
02071+3857   & 11.546   & 13.00 & 0.0179  &13496+0221   & 11.752   & 15.00 & 0.0328 \\
02203+3158   & 11.837   & 14.90 & 0.0338  &13536+1836   & 11.611   & 14.80 & 0.0497 \\
02222+2159   & 11.652   & 14.90 & 0.0338  &14151+2705   & 11.565   & 15.10 & 0.0365 \\
02248+2621   & 11.519   & 14.60 & 0.0327  &14178+4927*  & 11.541   & 15.40 & 0.0256 \\
02435+1253*  & 11.501   & 15.20 & 0.0216  &14547+2448*  & 11.897   & 14.60 & 0.0339 \\
02512+1446*  & 11.780   & 14.60 & 0.0312  &14568+4504   & 11.501   & 14.60 & 0.0357 \\
03117+4151   & 11.562   & 14.00 & 0.0235  &15107+0724*  & 11.525   & 15.50 & 0.0131 \\
(05084+7936) & 12.170   & 15.80 & 0.0543  &15163+4255*  & 12.072   & 14.90 & 0.0402 \\
(05414+5840) & 11.505   &  --   & 0.0148  &15327+2340*  & 12.464   & 14.40 & 0.0182 \\
(06538+4628) & 11.490   & 13.70 & 0.0214  &15425+4114   & 11.515   & 14.20 & 0.0317 \\
(07062+2041) & 11.559   &  --   & 0.0174  &15426+4116   & 11.546   & 13.90 & 0.0319 \\
07063+2043   & 11.570   & 12.60 & 0.0173  &16104+5235*  & 11.687   & 14.50 & 0.0294 \\
(07256+3355*)& 11.467   & 14.70 & 0.0135  &16180+3753   & 11.592   & 14.30 & 0.0307 \\
08354+2555*  & 11.781   & 14.40 & 0.0184  &16284+0411*  & 11.582   & 14.90 & 0.0246 \\
08507+3520   & 11.811   & 15.00 & 0.0559  &16504+0228*  & 12.028   & 14.70 & 0.0243 \\
(09047+1838) & 11.490   & 14.80 & 0.0291  &16577+5900*  & 11.582   & 14.20 & 0.0187 \\
09126+4432*  & 11.913   & 14.90 & 0.0393  &16589+0521   & 11.637   & 15.50 & 0.0502 \\
09168+3308   & 11.725   & 15.30 & 0.0499  &17366+8646   & 11.544   & 14.60 & 0.0264 \\
09320+6134*  & 12.220   & 15.50 & 0.0392  &17392+3845   & 11.554   & 15.00 & 0.0410 \\
09333+4841*  & 11.523   & 15.00 & 0.0259  &17501+6825   & 11.829   & 15.20 & 0.0512 \\
(09517+6954) & 10.833   & 9.57 &  0.0009  &18525+5518   & 11.683   & 15.50 & 0.0484 \\
10203+5235   & 11.620   & 15.00 & 0.0322  &18595+5048   & 11.501   & 15.10 & 0.0271 \\
(10311+3507) & 12.096   &  --   & 0.0710  &19120+7320   & 11.624   & 15.10 & 0.0250 \\
(10565+2448*)& 12.245   & 16.00 & 0.0431  &20550+1656*  & 12.074   & 15.20 & 0.0364 \\
11231+1456*  & 11.809   & 15.40 & 0.0341  &22388+3359*  & 11.531   & 15.00 & 0.0214 \\
11254+1126   & 11.800   & 14.80 & 0.0410  &22501+2427   & 11.723   & 15.30 & 0.0421 \\
11257+5850   & 12.040   & 11.80 & 0.0104  &23007+0836*  & 11.734   & 13.06 & 0.0162 \\
11543+0124   & 11.716   & 15.20 & 0.0397  &23024+1916*  & 11.573   & 15.20 & 0.0248 \\
(12112+0305) & 12.531   &  --   & 0.0724  &23135+2516*  & 11.730   & 15.00 & 0.0273 \\
12120+6838   & 12.029   & 15.40 & 0.0608  &23254+0830*  & 11.568   & 13.60 & 0.0290 \\
12251+4026   & 11.660   & 15.00 & 0.0371  &23488+1949*  & 11.528   & 13.39 & 0.0142 \\
12265+0219   & 12.663   & 13.07 & 0.1583  &23488+2018*  & 11.609   & 14.90 & 0.0179 \\
12323+1549   & 11.766   & 15.20 & 0.0461  &23532+2513   & 11.795   & 15.00 & 0.0571 \\
12540+5708*  & 12.636   & 14.10 & 0.0418  &23594+3622   & 11.586   & 15.40 & 0.0321 \\
12592+0436   & 11.787   & 15.50 & 0.0371  &             &          &       &       \\
\hline
\hline
\multicolumn{8}{l}{$^{a}$  The sources  marked * in column (1) were also observed by Kim et al.(1995)}\\
\end{tabular}
\end{table}
%END of Table 1
\end{document}